# Electrostatically driven helical plasma state

Cihan Akçay, John M. Finn, Richard A. Nebel, and Daniel C. Barnes

# Electrostatically driven helical plasma state

Cihan Akçay,[a)] John M. Finn, Richard A. Nebel, and Daniel C. Barnes
*Tibbar Plasma Technologies, 274 DP Rd., Los Alamos, New Mexico 87544, USA*

A novel plasma state has been found in the presence of a uniformly applied axial magnetic field in periodic cylindrical geometry. This state is driven electrostatically by helical electrodes, providing a driving field that depends on the radius and $m\theta - n\zeta$, where $\theta$ is the poloidal angle and $\zeta = z/R$ is the toroidal angle. We focus on $m = n = 1$. The radial magnetic field at the wall is taken to be zero. With weak driving, the resulting distortion is very small, but for stronger driving, the mean field of the state has field line safety factor $q_0(r)$ just above the pitch of the electrodes $m/n = 1$ except near the edge, where $q_0$ increases monotonically. This state is characterized as a single helicity Ohmic equilibrium with the helical symmetry of the applied field. The plasma appears to be close to force-free in the interior, but current density crosses the magnetic flux surfaces near the edge, where current must enter and exit through the helical electrodes. This perpendicular current density drives large helical plasma flows. The sensitivity of this state to flow boundary conditions, plasma resistivity profile, the strength of electrostatic driving, and parameters such as the loop voltage and the Lundquist number is explored. The magnetic helicity is calculated for both the transient period and time-asymptotic state. Possible applications to current drive in toroidal confinement devices and to electrical transformers are discussed. *Published by AIP Publishing.*
[http://dx.doi.org/10.1063/1.4981384]

## I. INTRODUCTION

Helical magnetic perturbations applied at the wall in toroidal plasma can be the basis of a plasma confinement scheme, as in stellarators, or can be seen as problematic, as in tokamaks. In the latter case, they are referred to as error fields although such applied perturbations can be useful as schemes to control instabilities.[20] It has also been suggested[1] that applying such helical magnetic perturbations to a reversed field pinch (RFP) could be useful in obtaining a robust quasi-single-helicity (QSH) state.[2,3] This scheme may have some merit, but there is one potential problem: in applying a magnetic perturbation, there will be a nonzero value of the normal component $B_n$ at the wall, allowing direct contact between the plasma and the wall along the magnetic field lines.

In this paper, we explore the idea of applying a helical *electrostatic* field at the radial wall of a plasma device,[1] arranged so that the normal component of the magnetic field is zero there. We present results in a periodic cylindrical geometry, representative of a model for a torus without toroidal geometric effects or a model for plasma of finite length but without end effects such as line tying, resistive walls, or sheaths. These studies are initialized with a uniform axial magnetic field $B_z = $ const., zero pressure, and azimuthal field $B_\theta = 0$. We apply a helical voltage at the wall ($r = r_w$), corresponding to physical electrodes, assuring at the same time $B_r(r_w) = 0$. In periodic cylindrical geometry, the voltage on the electrodes is of the form $\cos(m\theta + kz) = \cos(m\theta - n\zeta)$, where $\zeta = z/R$ is the toroidal angle, $R$ is the major radius, and $k = -n/R$.

In Sec. II, we introduce the zero $\beta$ resistive MHD model, in which the density is not advanced. The MHD equations are advanced with the DEBS[4] code and benchmarked with the NIMROD[5] code. The featured simulations employ a low Lundquist number $S$ since our immediate focus is on preliminary experiments in a small low-temperature device. We use a resistivity profile that is peaked near the edge to model sheaths around the electrodes. The applied helical electrostatic field is implemented as a potential of the form $\phi(r = r_w) = \phi_0 \cos(m\theta + kz)$, where $\phi_0$ corresponds to the electrode voltage. The choices for the plasma flow boundary conditions are also discussed.

In Sec. III, we show computational results for electrostatic driving with $(m,n) = (1, 1)$, indicating as expected that for a very small electrode voltage $\phi_0$, there is a weak perturbation. However, for a larger electrode voltage, the applied perturbation results in a helically symmetric state with $(m,n) = (1, 1)$, qualitatively different from the state with low $\phi_0$. Specifically, this state exhibits a significant poloidal mean field $B_\theta^{(0,0)}(r)$ that rises linearly with radius except near the edge, together with a nearly uniform axial mean field $B_z^{(0,0)}(r)$. As $\phi_0$ increases, the mean field safety factor $q_0(r) = rB_z^{(0,0)}(r)/RB_\theta^{(0,0)}(r)$ decreases monotonically from infinity and becomes very flat in the plasma interior as it approaches $m/n = 1$ from above in the regime of a large $\phi_0$. The mean current density $j_z^{(0,0)}(r)$ associated with $B_\theta^{(0,0)}(r)$ is approximately constant in the plasma interior but decreases and changes sign near the edge. Neither $j_z^{(0,0)}(r)$ nor $B_z^{(0,0)}(r)$ is nearly as flat as $q_0(r)$ in the interior for a large $\phi_0$. The helical flux contours, labeling flux surfaces in helical symmetry, resemble those of a model with a large $m = 1$ perturbation, well past a Kadomtsev reconnection.[6–8] These flux

---

[a)]Electronic mail: c_akcay@tibbartech.com

surfaces show that all of the field lines encircle an elliptic field line (*O*-point) near the edge of the plasma, similar to the single helical axis (SHAx) state in RFPs.[9,10] They also indicate that the effect of the perturbation is very strong in the plasma interior, even with a relatively low applied helical field, because this region has a weak shear $dq_0/dr$ and $q_0 \gtrsim m/n = 1$. The *helical field*, whose surfaces correspond to the current density lines in helical symmetry, aligns with the helical flux surfaces to some degree in the interior where the plasma is nearly force-free and $q_0$ is nearly flat. However, the contours of the helical flux and helical field differ significantly at the edge, where the current from the electrodes crosses the field lines, requiring $\mathbf{j} \times \mathbf{B} \neq 0$. In this boundary region, the Lorentz force $\mathbf{j} \times \mathbf{B}$ drives strong helical flows, determined by a balance between $\mathbf{j} \times \mathbf{B}$ and either the inertia or the viscous force.

Above a certain value of $\phi_0$, the steady-state equilibrium with the characteristics described above is not observed in the simulations, and a time-dependent state appears. This transition is associated with the value of the ratio of the norm of $\mathbf{v}_\perp$ to the (nearly constant) Alfvén speed. Specifically, this occurs when this ratio, the Alfvén Mach number, approaches unity.

According to the parallel Ohm's law, the flux surface average $\langle \eta \lambda B^2 \rangle$ must be zero on all of the flux surfaces in steady state, where $\lambda = \mathbf{j} \cdot \mathbf{B}/B^2 \equiv j_\parallel/B$. The condition $\eta B^2 > 0$ dictates a change in sign of $\lambda$ along the field line, i.e., the parallel current density described by $\lambda$ consists of Pfirsch-Schlüter currents[11] associated with the flow instead of pressure. The parallel current density $\lambda$ is zero along the *O*-lines, and this property, not present for $\theta$ or $z$ symmetry by Cowling-like theorems,[2,12] is allowed by the helical geometry, i.e., by stellarator transform.[2]

In Sec. IV, we investigate the effects of (A) specifying a helical current density $j_r$ at the wall instead of the applied potential; (B) employing a uniform (flat) resistivity profile instead of one that is peaked at the edge; (C) varying the boundary conditions on the velocity; (D) applying an axial loop voltage (back EMF) to simulate the effect of a secondary circuit; and last (E) increasing the Lundquist number $S$. The characteristics of the nominal state persist into all these regimes, undergoing only a few notable changes. For a flat resistivity profile, the negative values of $\lambda$ in the flux surface average are more pronounced than that for the nominal case with a peaked resistivity profile, as expected from $\langle \eta \lambda B^2 \rangle = 0$. In the toroidal context, the axial voltage represents the inductive loop voltage applied to drive the axial current. For the finite length system, this potential represents the back EMF due to driving current through a circuit with a load. This EMF has the opposite sign as the loop voltage that drives a positive current in the toroidal case. Increasing $S$ leads to qualitatively similar results, with a sharper transition to the regime with $q_0(r) \approx m/n = 1$ and with this regime spanning a wider range of helical voltage before the non-steady state is encountered.

The helical nature of the electrostatic drive and the plasma quantities that emerge as a result naturally raise questions regarding magnetic helicity. In Sec. V, we present a discussion of the magnetic helicity for both the transient period and time-asymptotic state, without and with the back EMF.

In Sec. VI, we discuss two possible applications of this novel plasma state driven by helical electrodes. In Sec. VII, we summarize and discuss our results. The Appendix describes the single helicity quantities used as code diagnostics as well as the effect of a constant density assumption on energy conservation in MHD.

## II. THE COMPUTATIONAL MODEL

We use the zero-pressure resistive MHD equations with a constant and uniform density. These equations comprise the equation of motion, the resistive Ohm's law, and Faraday's law

$$\rho_0 \left( \frac{\partial \mathbf{v}}{\partial t} + \mathbf{v} \cdot \nabla \mathbf{v} \right) = \mathbf{j} \times \mathbf{B} + \mu \nabla^2 \mathbf{v}, \quad (1)$$

$$\mathbf{E} + \mathbf{v} \times \mathbf{B} = \eta \mathbf{j}, \quad (2)$$

$$\frac{\partial \mathbf{B}}{\partial t} = -\nabla \times \mathbf{E}, \quad (3)$$

where $\mathbf{B}$ and $\mathbf{E}$ are the magnetic and electric fields, $\mathbf{j} = \nabla \times \mathbf{B}$ the current density, $\rho_0$ the (constant) plasma density, and last $\mathbf{v}$ the center-of-mass plasma flow velocity. The quantities $\eta$ and $\mu$ are the plasma resistivity and viscosity, respectively. The geometry is a periodic cylinder, occupying $0 \leq r \leq r_w$ and $0 \leq z \leq L = 2\pi R$.

The above equations have been non-dimensionalized by scaling lengths to the wall radius, $r_w = 1$, the initial magnetic field $\mathbf{B}$ to the initial constant value $B_z = B_0$, and time to the nominal Alfvén time $\tau_A = r_w/v_A$, where the Alfvén speed is based on $B_0$ and $\rho_0 = 1$. The velocity in these units is then relative to $v_A$. The plasma resistivity can have a spatial variation: $\eta = \eta(r)$, while the viscosity, $\mu$, is kept spatially uniform. The Lundquist number is defined as $S = \tau_R/\tau_A$, where $\tau_R = r_w^2/\eta(r=0)$ is the resistive diffusion time. (DEBS rescales time in terms of $\tau_R$, which introduces factors of $S$ into Eqs. (1)–(3).[4]) The Reynolds number is $Re = \tau_v/\tau_A = 1/\mu$, where $\tau_v = r_w^2/\nu$ is the viscous diffusion time, where $\nu$ is the kinematic viscosity. The magnetic Prandtl number is $Pr = \mu/\rho_0 \eta(0) = S/Re$. As indicated above, the density is not advanced, since an accurate treatment would involve ionization, recombination, and other phenomena present in low-temperature plasmas. Furthermore, MHD phenomena do not depend sensitively on the density profile. As discussed in the Appendix and in Ref. 13, the assumption $\rho = \rho_0 = $ const. violates the conservation of energy in a dissipationless system ($\eta = \mu = 0$), although we argue that this effect is weak for these simulations.

The resistivity is specified as a function of radius to be of the form

$$\eta(r) = \eta(0) \left[ 1 + (\sqrt{\eta(1)/\eta(0)} - 1) r^p \right]^{1/2} \quad (4)$$

for $p = 16$. For $\eta(1) \gg \eta(0)$, this prescription provides a large resistivity near the wall, with $\eta(r) \approx \eta(0)$ in the interior. The high edge resistivity is a model for sheaths around electrodes. A case with uniform resistivity profile $\eta(1)/\eta(0) = 1$ is also presented in Sec. IV B.

Equations (1)–(3) are advanced with the DEBS code[4] and have been benchmarked by the NIMROD code.[5] The focus of the present work is on the results from the DEBS simulations. DEBS advances the vector potential $\mathbf{A}(\mathbf{x}, t)$ rather than the magnetic field $\mathbf{B}(\mathbf{x}, t)$. It uses the Weyl or temporal gauge,[14] $\phi = 0$, which results in $\mathbf{E} = -\partial \mathbf{A}/\partial t$. For the spatial discretization, DEBS uses a finite difference approximation for the radial variation and a Fourier representation for the variation of the fields in the poloidal and axial directions, e.g., $\mathbf{E}(\mathbf{r}) = \mathbf{E}(r)e^{im\theta + ikz} = \mathbf{E}(r)e^{im\theta - in\zeta}$, where the toroidal angle is $\zeta = z/R$, $R$ is the major radius, and $k = -n/R$. For the remainder of this paper, we focus on $m = n = 1$.

The boundary conditions on the fields are as follows: for voltage boundary conditions at the wall, we specify a voltage $\phi_0$ at $r = r_w = 1$ with the two conditions

$$E_\theta^{(m,n)}(1) = -(im/r_w)\phi_0 e^{im\theta + ikz} = -i\phi_0 e^{i\theta - i\zeta}, \text{ for } m = n = 1 \tag{5}$$

and

$$E_z^{(m,n)}(1) = -ik\phi_0 e^{im\theta + ikz} = i(\phi_0/R)e^{i\theta - i\zeta}, \text{ for } m = n = 1, \tag{6}$$

with $E_\theta^{(-1,-1)}(1) = E_\theta^{(1,1)*}(1)$ and similarly for $E_z$. Also, we take the tangential components $E_\theta^{(m,n)} = E_z^{(m,n)} = 0$ for all other $(m, n)$. These relations result in

$$mE_z^{(1,1)} - krE_\theta^{(1,1)} = E_z^{(1,1)} + \frac{r}{R}E_\theta^{(1,1)} = 0. \tag{7}$$

These relations are consistent with $(\partial/\partial t)B_r = 0$. (We also assume $B_r(t = 0) = 0$.) Here, we have the helical coordinates $\hat{\mathbf{r}}$, $\mathbf{k} = \nabla u$ with $u = m\theta + kz = m\theta - n\zeta$, and $\boldsymbol{\sigma} = \hat{\mathbf{r}} \times \mathbf{k}$. Eq. (7) is equivalent to $r\boldsymbol{\sigma} \cdot \mathbf{E} = 0$. This condition allows a tangential component of $\mathbf{E}$ at the wall that is parallel to $\mathbf{k}$, $\mathbf{E}_t = -\nabla_t \phi_0 e^{im\theta - in\zeta}$, as in Eqs. (5) and (6). Since DEBS uses the Weyl gauge, the actual boundary conditions applied to the vector potential $\mathbf{A}(\mathbf{x}, t)$ are $A_\theta^{(1,1)}(1) = (im/r_w)t\phi_0 e^{iu}$ and $A_z^{(1,1)}(1) = ik\phi_0 t e^{iu}$. Note that $A_r$ does not need to be specified at the wall because resistive diffusion applies only to the transverse part of $\mathbf{A}$. The condition in Eq. (7) is weaker than the perfectly conducting condition $E_\theta^{(1,1)} = E_z^{(1,1)} = 0$.

We also explore an alternate formulation of the EM boundary conditions by specifying $j_r^{(1,1)}$ (and of course $j_r^{(-1,-1)}$) at $r = 1$, along with the condition (7) for $(m,n) = (1, 1)$. See Sec. IV A. For all other $(m, n)$, the tangential electric field is set to zero. For the first two cases of velocity boundary conditions, with tangential components of $\mathbf{v}$ set to zero at the wall, we have $E_r^{(1,1)} = \eta j_r^{(1,1)}$ at the wall; in this case, the boundary condition on $j_r^{(1,1)}$ is also an inhomogeneous boundary condition on $E_r^{(1,1)}$. In Sec. IV A, we describe a comparison of these two EM boundary conditions, namely, specifying the tangential components of $\mathbf{E}$ for $(m,n) = (1, 1)$ as in Eqs. (5) and (6) with the tangential components of every other Fourier component equal to zero; and specifying $j_r^{(1,1)}$, $j_r^{(-1,-1)}$ with the tangential components of $\mathbf{E}$ for all other values of $(m, n)$ equal to zero. For these two specifications, a successful benchmarking test is discussed. The effect of the broadening of the spectrum, especially for more strongly driven cases, is also discussed.

Three alternate sets of boundary conditions on the plasma velocity at $r = r_w$ are imposed: (I) $v_r(1) = (\mathbf{E} \times \mathbf{B}/B^2)_r$ for the Fourier amplitudes with $(m, n) = (0, 0)$, $(1, 1)$ (and $(-1, -1)$), while the remaining velocity components are set to zero, (II) homogeneous Dirichlet conditions, i.e., all three components of the plasma velocity for all Fourier amplitudes are set to zero, and (III) homogeneous Neumann boundary conditions applied to all three components of all Fourier amplitudes, i.e., $\partial \mathbf{v}(1)/\partial r = 0$. For the first set of boundary conditions, only the product of the mean $\mathbf{E}$ and $\mathbf{B}$ are used for $v_r^{(0,0)}(1)$: $v_r^{(0,0)}(1) = \hat{\mathbf{r}} \cdot \mathbf{E}^{(0,0)} \times \mathbf{B}^{(0,0)}/(B^{(0,0)})^2$, neglecting small quasilinear (convolution) terms like $\hat{\mathbf{r}} \cdot \mathbf{E}^{(1,1)} \times \mathbf{B}^{(-1,-1)}/(B^{(0,0)})^2$. The field $\mathbf{E}^{(0,0)} = E_0 \hat{\mathbf{z}}$ is the back EMF discussed in Sec. IV D and is zero for the nominal case studied in Sec. III. For the (1, 1) component, we take $v_r^{(1,1)}(1) = \hat{\mathbf{r}} \cdot \mathbf{E}^{(1,1)} \times \mathbf{B}^{(0,0)}/(B^{(0,0)})^2$, linearizing with respect to the applied (1, 1) electric field, and similarly for $(m,n) = (-1, -1)$. The homogeneous Dirichlet condition at the wall does not allow plasma flux in or out of the plasma. This does not pose a problem with the density $\rho$ because, as we have noted, $\rho$ is not evolved. These three different prescriptions for the velocity allow us to test the sensitivity to the velocity boundary conditions in Sec. IV C. Note that for the first two prescriptions, having the tangential components of the velocity zero, the radial electric field at the wall is $E_r(1) = \eta(1)j_r(1)$.

### III. CHARACTERIZATION OF THE NOMINAL STATE

The first case, which is referred to as the "nominal" case, has $R = 3$ (aspect ratio of 3), $S = 100$, $Pr = 10$, and $\eta(1)/\eta(0) = 100$. The helical perturbation is applied as an electrostatic potential consistent with the electric fields of Eqs. (5) and (6) for various values of $\phi_0$. As will be shown in Sec. IV A, prescribing the current at the boundary $j_r^{(1,1)}(r = 1)$ to apply the helical perturbation yields identical results to specifying $\phi_0$. For this nominal case, the first set of boundary conditions on the velocity are applied, that is, $v_r^{(1,1)} = (E_\theta^{(1,1)} B_z^{(0,0)} - E_z^{(1,1)} B_\theta^{(0,0)})/(B^{(0,0)})^2$. We have $v_r^{(0,0)} = 0$ because there is no loop voltage for this case, i.e., $E_z^{(0,0)} = 0$, and $E_\theta^{(0,0)}$ is zero for all cases. (See Sec. IV D for nonzero back EMF.)

Here in Sections III–VIII, for diagnostic purposes, we will switch from the Weyl gauge used in DEBS to the Coulomb gauge $\nabla \cdot \mathbf{A} = 0$, in which the potentials have a clearer physical meaning. In the results shown in this section, the time asymptotic state is a steady state, achieved in the simulations in a few resistive decay times. In the Coulomb gauge, we have $\mathbf{E} = -\nabla \phi$ in steady state, with $\partial \mathbf{A}/\partial t = 0$. In contrast, in the Weyl gauge, with $\phi = 0$, the vector potential for steady-state fields is $\mathbf{A}(\mathbf{x}, t) = \mathbf{A}_0(\mathbf{x}) - t\mathbf{E}(\mathbf{x})$, with $\nabla \times \mathbf{E} = 0$, giving $\mathbf{B} = \nabla \times \mathbf{A}_0$.

## A. Properties of the nominal case

For a very small applied potential $\phi_0$, computations show that the steady state has small perturbed electric and magnetic fields. A simple linear calculation, similar to that of Refs. 15–20 for magnetic field errors, can be employed. This calculation, linearizing in $\phi_0$, indicates small perturbed fields in the plasma volume. The (1, 1) perturbation is very stable because it has $F = \mathbf{k} \cdot \mathbf{B} = kB_0 = -B_0/R$ quite large, leading to strong field line bending and suppression of the perturbation.

As the potential $\phi_0$ increases, the steady-state solution changes continuously to a state with a significant mean current density $j_z^{(0,0)}(r)$. This final state consists of (1, 1) perturbed fields driven by the helical boundary conditions and a mean field $B_\theta^{(0,0)}$ driven quasilinearly by the (1, 1) distortion. The simulations show that it is still single-helicity in form, including higher $(m, n)$, but with the same helicity $m/n = 1$ as the driving electric field. That is, the fields and plasma velocity are two-dimensional, behaving as $\mathbf{E}(r, \theta, z) = \mathbf{E}(r, u)$. Other Fourier harmonics with $m/n \neq 1$ are observed to be many orders of magnitude smaller.

The mean fields $B_\theta^{(0,0)}(r)$ and $B_z^{(0,0)}(r)$ lead to a profile of "mean" safety factor

$$q_0(r) = \frac{rB_z^{(0,0)}(r)}{RB_\theta^{(0,0)}(r)} \quad (8)$$

that increases across the plasma. The $q_0(r)$ profiles for four values of $\phi_0$ are shown in Fig. 1. The traces are normalized by the value of $q_0(r)$ on axis: $q_0(0)$. As $\phi_0$ increases, $q_0(0)$ approaches unity from above and the profile of $q_0(r)$ becomes much flatter for $r \lesssim r_1 = 0.7$ ($q_0(0)$ as a function of $\phi_0$ is plotted in Fig. 3). The ratio $q_0(1)/q_0(0)$ begins to decrease noticeably for $\phi_0 \gtrsim 0.2$. It is noteworthy that for large values of $\phi_0$, $B_z^{(0,0)}(r)$ and $B_\theta^{(0,0)}(r)/r$ both decrease significantly for $r < r_1$ while $q_0(r) \propto B_z^{(0,0)}/(B_\theta^{(0,0)}/r)$ (for small $r$) remains much flatter than either of these in this region. The flat $q_0(r)$ profile for larger $\phi_0$ within $r < r_1$ is a defining characteristic of this saturated state and persists over a wide range of parameters, as will be shown in Sec. IV. The closeness of $q_0$ to unity in the interior with very weak shear provides a strong sensitivity to the driven $(m, n) = (1, 1)$ helical distortion, which has a relatively weak field in that region. In other words, the flux surfaces are strongly distorted in this region, and the driven (1, 1) state and the low shear region develop together. The magnitude of the perturbed field is $||\mathbf{B}^{(1,1)}|| \sim 0.03$, where $||\mathbf{a}|| \equiv \int |\mathbf{a}| dV / \pi r_w^2 L$ ($L = 2\pi R$) is a volume average of the 2-norm of any vector $\mathbf{a}$. Notice also that $q_0(r) > 0$ implies the same sign for the pitch of the field lines ($\propto 1/q_0$) as $n/m = 1$ and the same sign for $j_z^{(0,0)}(r = 0)$ as $B_z^{(0,0)}$. Furthermore, letting $\phi_0 \to -\phi_0$ does not change this relationship because this transformation is equivalent to $\theta \to \theta + \pi$.

Contours of several important quantities are shown in Fig. 2. The helical flux (Fig. 2(a)) is $\chi(r, u) = r\boldsymbol{\sigma} \cdot \mathbf{A}$, or

$$\chi(r, u) = mA_z(r, u) - krA_\theta(r, u) = A_z(r, u) + rA_\theta(r, u)/R, \quad (9)$$

where $m = n = 1$ is substituted into the latter form. This quantity is gauge invariant and proportional to the magnetic flux through a ribbon perpendicular to $u = \text{const.}$. The conditions $B_r = (1/r)\partial\chi/\partial u$ and $\mathbf{k} \cdot \mathbf{B} = -(1/r)(\partial\chi/\partial r)$ imply $\mathbf{B} \cdot \nabla\chi = B_r(\partial\chi/\partial r) + \mathbf{B} \cdot \nabla u\, (\partial\chi/\partial u) = 0$, meaning that $\chi$ surfaces are magnetic surfaces (see the Appendix). The $\chi$ contours shown in Fig. 2(a) have no separatrix, consistent with $q_0(r) > 1$. They exhibit strong (1, 1) structure in spite of a relatively small (1, 1) field ($||B_\theta^{(1,1)}||/||B_\theta^{(0,0)}|| = 0.1$), because of the flat $q_0(r) \gtrsim 1.0$, as discussed above. Because of the dominant $m = 1$ structure of the plasma flow, the (elliptic) O-point is strongly displaced to one side of the domain. The reader can refer to the Appendix for further discussion on $\chi$ as well as the other helical quantities presented below.

The axial current density $j_z$ shown in Fig. 2(b) indicates a bipolar current structure, as does the parallel current density $\lambda = \mathbf{j} \cdot \mathbf{B}/B^2$, in Fig. 2(d). The two quantities are very similar because $B_z$ is nearly uniform and the current is nearly force-free in the interior, resulting in $j_z \approx \lambda B_z \approx \lambda$. The thick curve in Figs. 2(a)–2(d) and 2(f) corresponds to $\lambda = 0$. The curve representing $j_z = 0$ is virtually indistinguishable. Note that the O-point in the $\chi$ contours intersects the $\lambda = 0$ contour. We will return to this issue in Sec. III B. Furthermore, note that the position of the O-point coincides with the radius where the shear and the value of $q_0 - 1$ in Fig. 1 begin to increase significantly. Also, in Fig. 2(b), the positively directed current density occupies a significantly larger area but it is smaller in magnitude, and the same holds for $\lambda > 0$. The total current $I_z = \int j_z r dr d\theta$ is positive. We will return to this issue in Secs. IV B and IV D.

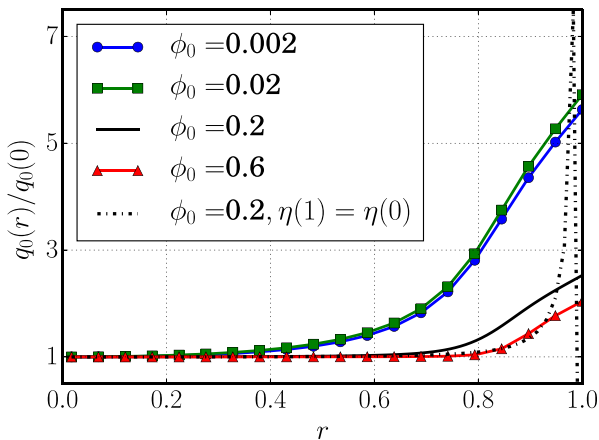

FIG. 1. The mean field profile $q_0(r)$ normalized by $q_0(r = 0)$ for the nominal case driven with four different values of the applied potential: $\phi_0 = 0.002, 0.02, 0.2,$ and $0.6$ for a $(m, n) = (1, 1)$ electrostatic perturbation. Also shown is an additional case for $\phi_0 = 0.2$ with uniform resistivity, i.e., $\eta(r = 1) = \eta(r = 0)$, as a dashed-dotted black line (discussed in Sec. IV B). The mean safety factor blows up in this last case because $B_\theta^{(0,0)} \approx 0$ near the wall. The actual values of $q_0(r)$ for small $\phi_0$ are orders of magnitude larger than 1.0, e.g., $q_0(0) = 545$ for $\phi_0 = 0.002$.

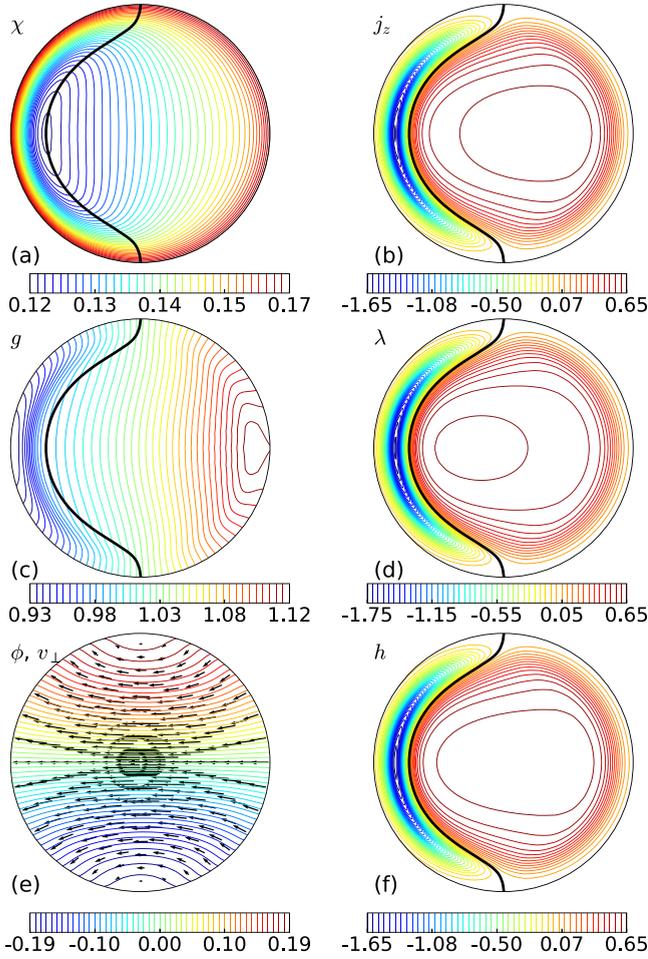

FIG. 2. Contours for salient physical quantities from the nominal case ($\phi_0 = 0.2$) at $z = 0.5L$ (a) helical flux $\chi$, (b) axial current density $j_z$, which indicates a bipolar current flow, (c) helical field $g$, whose contours are mostly aligned with those of $\chi$ in the interior but show current density entering and exiting at $r = 1$, (d) parallel current $\lambda \equiv \mathbf{j} \cdot \mathbf{B}/B^2$, (e) electrostatic potential $\phi$ with vectors of $\mathbf{v}_\perp$, and (f) $h = r\boldsymbol{\sigma} \cdot \mathbf{j}$. The three quantities $\lambda$, $h$, and $j_z$ are all similar since the current is nearly force-free in the interior, resulting in $j_z \approx \lambda B_z$, and $B_z$ is nearly uniform. The black contours in (a)–(d), and (f) correspond to $\lambda = 0$.

Analogous to the helical flux, the helical *field*

$$g(r, u) = r\boldsymbol{\sigma} \cdot \mathbf{B} = mB_z - krB_\theta = B_z + rB_\theta/R \quad (10)$$

satisfies $\mathbf{j} \cdot \nabla g = 0$, implying that current density streamlines lie on $g(r, u)$ surfaces. Note that since $rB_\theta/(RB_z) \ll 1$, $g \simeq B_z$. The current density satisfies $\boldsymbol{\sigma} \cdot \mathbf{j} \times \mathbf{B} \propto \boldsymbol{\sigma} \cdot \nabla \chi \times \nabla g$, so that the $g$ surfaces and the $\chi$ surfaces should approximately align where the plasma is nearly force-free. Figure 2(c) displays $g$ surfaces for the nominal case, showing that the field lines and current density lines do align fairly well except near the wall, where the lines of $g$ must enter the electrodes, where $j_r = (1/r)\partial g/\partial u$. The structure of $g$ is consistent with the applied potential and resulting current flow across the plasma. Also apparent in Fig. 2(c) is a small area to the right with a separatrix, which contains closed current surfaces that are not connected to the helical electrodes. We revisit this issue in Secs. III C and IV B, where we present other cases with a larger amount of current within this separatrix (or no $g$ separatrix at all).

The perpendicular flow $\mathbf{v}_\perp$ along with the contours of the electrostatic potential (in the Coulomb gauge—see the Appendix) are shown in Fig. 2(e); $\mathbf{v}_\perp$ aligns well with the $\phi$ surfaces except near the boundary. This is consistent with Ohm's law, Eq. (2), which shows

$$\mathbf{v}_\perp = \frac{\mathbf{B} \times (\nabla \phi + \eta \mathbf{j})}{B^2}. \quad (11)$$

If $\eta|\mathbf{j}_\perp| \ll \phi/r_w$, the velocity is the $E \times B$ drift, $\mathbf{v}_\perp \approx \mathbf{v}_{E \times B}$. By comparing Figs. 2(a) and 2(e), it is evident that advection of the flux by $\mathbf{v}_\perp$ sustains its strongly distorted $(m,n) = (1, 1)$ character. The volume-averaged 2-norm of perpendicular velocity, $\|\mathbf{v}_\perp\|$—defined previously—reaches 0.2 for this case. Figure 2(e) indicates that $\nabla \cdot \mathbf{v}_\perp$ is relatively large near the wall at the right and the left, related to the value of $v_r$ through the wall and related to small return flows at the edge. Indeed, for this nominal case $|\nabla \cdot \mathbf{v}_\perp|$ and $|\nabla \cdot \mathbf{v}_{E \times B}|$, both relative to $|\mathbf{v}_\perp|/r_w$, are of order unity in these regions (see the Appendix).

The component of $\mathbf{j}$ along $\boldsymbol{\sigma}$, $h = r\boldsymbol{\sigma} \cdot \mathbf{j}$, is shown in Fig. 2(f). This quantity emerges from the representation of the current density based on the helical quantities described thus far: $\mathbf{j} = f(r)\nabla g \times \boldsymbol{\sigma} + f(r)h\boldsymbol{\sigma}$ (see the Appendix). Note that $h \simeq j_z$, based on the same arguments that $g \simeq B_z$; indeed, $h$ exhibits a structure very similar to that seen in $j_z$ and $\lambda$.

The magnitude of the Lorentz force $\mathbf{j} \times \mathbf{B}$, normalized to the maximum of $|\mathbf{j}||\mathbf{B}|$ (not shown), is fairly small, but peaks near the wall, with a value of 0.08, in the vicinity of the O-line in the $\chi$ surfaces. Since the field lines are tangential at the wall, the electrode current density crosses field lines in that vicinity, as indicated in Fig. 2(c). In steady state (with $\beta = 0$), the Lorentz force in Eq. (1) must be balanced by either inertia or viscosity. In either case, the Lorentz force drives helical flows.

The helical nature of the applied potential is essential for driving a mean axial current density $j_z^{(0,0)}$. Consider an $(m, n) = (1, 0)$ distortion in lieu of the $(1, 1)$ used in this work. The principle that the mean fields $B_\theta^{(0,0)}$ and $B_z^{(0,0)}$ line up with the electrode pitch, i.e., that $\mathbf{k} \cdot \mathbf{B} = B_\theta^{(0,0)}/r \to 0$ for large $\phi_0$ leads to $B_\theta^{(0,0)} \to 0$. This implies $A_z^{(0,0)} = 0$ as well as $\chi^{(0,0)} = mA_z^{(0,0)} = 0$ ($n = 0$ here). These predictions are in agreement with the results of a simulation that was performed with a potential having $(m, n) = (1, 0)-$ symmetry: the mean fields indeed have $B_\theta^{(0,0)} = 0$, with $B_z^{(0,0)}$ varying slightly near the wall.

### B. Flux surface average condition in steady state

The steady state solution shown in Figs. 1 and 2 has $\mathbf{E} = -\nabla \phi$ in the Coulomb gauge. The parallel component of Ohm's law, Eq. (2), is

$$-\mathbf{B} \cdot \nabla \phi = \eta \mathbf{j} \cdot \mathbf{B} = \eta \lambda B^2. \quad (12)$$

Since the flux surface average of the left side of Eq. (12) is zero (see the Appendix), we find

$$\langle \eta \lambda B^2 \rangle = 0. \quad (13)$$

Any flux surface in a hypothetical force-free region, where $\mathbf{B} \cdot \nabla \lambda$ is exactly zero, has $\lambda(\chi) \langle \eta B^2 \rangle = 0$, implying $\lambda = 0$ since $\langle \eta B^2 \rangle > 0$ everywhere. However, as seen in Figs. 2(a) and 2(d), the flux surfaces all connect the region with $\lambda > 0$ to the reversal region where $\lambda < 0$. Conversely, the condition $\mathbf{B} \cdot \nabla \lambda \neq 0$ implies $\mathbf{j} \times \mathbf{B} \neq 0$, since $\nabla \cdot \mathbf{j}_\perp = -\mathbf{B} \cdot \nabla \lambda$. These arguments show that the parallel current density consists of Pfirsch-Schlüter currents,[7,11] here associated with the flow rather than with pressure.

Note also that there is an elliptic field line (O-line) at the center of the flux surfaces in Fig. 2(a). This field line must follow the $(m,n)=(1,1)$ helical twist exactly, implying the following from Eq. (12): $-B(d\phi/dl) = \eta \lambda B^2$, so $\Delta\phi = -\oint \eta \lambda B^2 (dl/B)$, where the integration is along the whole closed field line. Since $\Delta\phi = 0$, we have the analog of Eq. (13), namely,

$$\oint \eta \lambda B dl = 0. \quad (14)$$

By symmetry, all quantities must be constant along the O-line. This leads to $\oint \eta \lambda B dl = \eta \lambda B L_O = 0$, where $L_O$ is the length of the O-line. This equality is only satisfied if $\lambda = 0$ on this field line, in agreement with the earlier observation from Fig. 2(a) that the $\lambda = 0$ contour passes exactly through the O-point of the helical flux. Thus, the helical symmetry allows an O-line with zero parallel current density, $\lambda = 0$, which is forbidden in axisymmetry. That is, the nested flux surfaces around the O-line are due to helical (stellarator) transform rather than local current density. The helical nature of the geometry, namely, $\boldsymbol{\sigma} \cdot \nabla \times \boldsymbol{\sigma} \propto mk \neq 0$, makes this effect possible.[2] As discussed in Refs. 21 and 22, this effect is responsible for the possibility of dynamo-like behavior in spheromaks with good flux surfaces.

## C. Variation of driving voltage

It is of interest to determine how certain physical quantities that serve as metrics vary as functions of the driving voltage. The quantities in question here are plotted in Fig. 3 over a range in the applied voltage $\phi_0$ spanning $[2 \times 10^{-5}, 0.7]$. They are the reciprocal of $q_0$ on the axis ($r=0$), $1/q_0(0)$ (solid blue), total plasma current $I_z$ (green dots), volume average of the 2-norm of the perpendicular velocity $\|\mathbf{v}_\perp\|$ (red arrows), and the radial position of the O-point in $\chi$, $r_O$ (cyan with squares). Recall $1/q_0(0) \propto j_z^{(0,0)}(0)$. The current $I_z$ is scaled by $B_0 r_w/\mu_0$ in this dimensionless units. For $B_0 = 0.1$ T and $r_w = 0.1$ m, $I_z = 1.0$ is equivalent to approximately 8 kA of current.

In the range $\phi_0 < 0.02$, there is little mean current driven, i.e., $q_0(0)$ is large while $I_z$ and $\|\mathbf{v}_\perp\|$ remain small and the flux surfaces are centered at the origin, i.e., $r_O \approx 0$. Up to $\phi_0 \approx 0.05$, the profile $q_0(r)/q_0(0)$ appears to rise modestly in the interior, as illustrated in Fig. 1. Also, $1/q_0(0) \propto \phi_0^2$, due to the quasilinear generation of $B_\theta^{(0,0)}$ by the $(m,n)=(1,1)$ (and $(-1,-1)$) terms. This scaling breaks down for larger $\phi_0$, leading to $1/q_0(0) \to 1$. Up to $\phi_0 = 0.7$, a steady state still occurs, with $q_0(0) \gtrsim 1.0$ while the whole $q_0(r)$ profile continues to get increasingly flatter, as depicted by Fig. 1. The (perpendicular) flow appears to rise linearly in the moderate-to-strongly driven regime. In fact, $\|\mathbf{v}_\perp\| \approx \phi_0$. This is a consequence of $\mathbf{v}_\perp \approx \mathbf{v}_{E \times B}$ and $\mathbf{v}_{E \times B}/v_A \approx (\phi_0/r)/(B_0 v_A)$, which upon carrying out the volume integration yields $\|\mathbf{v}_{E \times B}\| \approx \phi_0$ for $r_w = B_0 = v_A = 1$. In other words, $\phi_0$ is an accurate measure of the average flow speeds relative to the Alfvén speed in the simulations. At large values of $\phi_0$ for which $q_0(0)$ becomes very close to unity, the flow becomes Alfvénic. Above $\phi_0 = 0.7$, the steady single-helicity state is not observed, and rather the fields and flow have irregular time dependence.

Figure 4 shows contours of $\chi$ (left column) and $\lambda \approx j_z$ (right column) for three different values of the applied potential: $\phi_0 = 2 \times 10^{-3}$, 0.02, and 0.6 from top to bottom, corresponding to steady state solutions for weak, intermediate, and strong driving, respectively. For the first case, which has $q_0(0) = 545$, the helical flux contours show little indication of a helical perturbation. Since $j_z^{(0,0)}$ is so small (note $\lambda \approx j_z^{(1,1)}$ in Fig. 4(b)), the helical flux satisfies $\chi^{(0,0)} \approx -kr A_\theta^{(0,0)} \approx B_0 r^2/2R$. That is, $q_0(0)$ is so large that the (1, 1) perturbation is very far off-resonance, and $\chi \approx \chi^{(0,0)}$. For the second case, with $q_0(0) \approx 5$, the $\chi$ contours (Fig. 4(c)) show somewhat more helical distortion and the $\lambda$ contours (Fig. 4(d)) show some asymmetry due to the larger value of $j_z^{(0,0)}$. In the third case, with $q_0(0) = 1.01$, the $\chi$ and $\lambda$ contours (Figs. 4(e) and 4(f)) exhibit considerable helical distortion, both because the helical perturbation is larger but also because $q_0$ is extremely close to unity over a large range in radius.

## IV. VARIATIONS ON THE NOMINAL CASE

In this section, we explore the robustness of the main results of Sec. III to various modifications. These are (1) the specification of the helical perturbation in terms of the normal current density at the wall $j_r(r=1)$ rather than the potential $\phi_0$; (2) simulating a plasma with a uniform resistivity profile in lieu of the non-uniform resistivity employed in Sec. III; (3) imposing zero-flow (homogeneous Dirichlet) plasma velocity boundary conditions or, alternatively, homogeneous

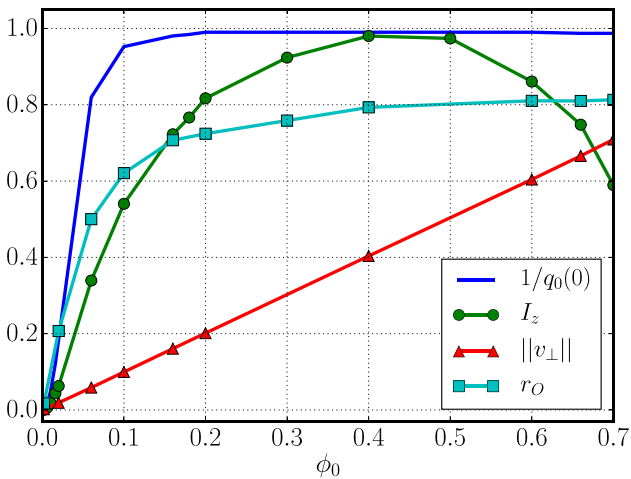

FIG. 3. Inverse $q_0(0)$ (solid blue), total axial current $I_z$ (green), volume average of the 2-norm of $\mathbf{v}_\perp$ (red) defined as $\|\mathbf{v}_\perp\| \equiv \int |\mathbf{v}_\perp| dV/\pi r_w^2 L$, and radial position $r_O$ of the O-point (cyan) as functions of the amplitude of the applied voltage $\phi_0$.

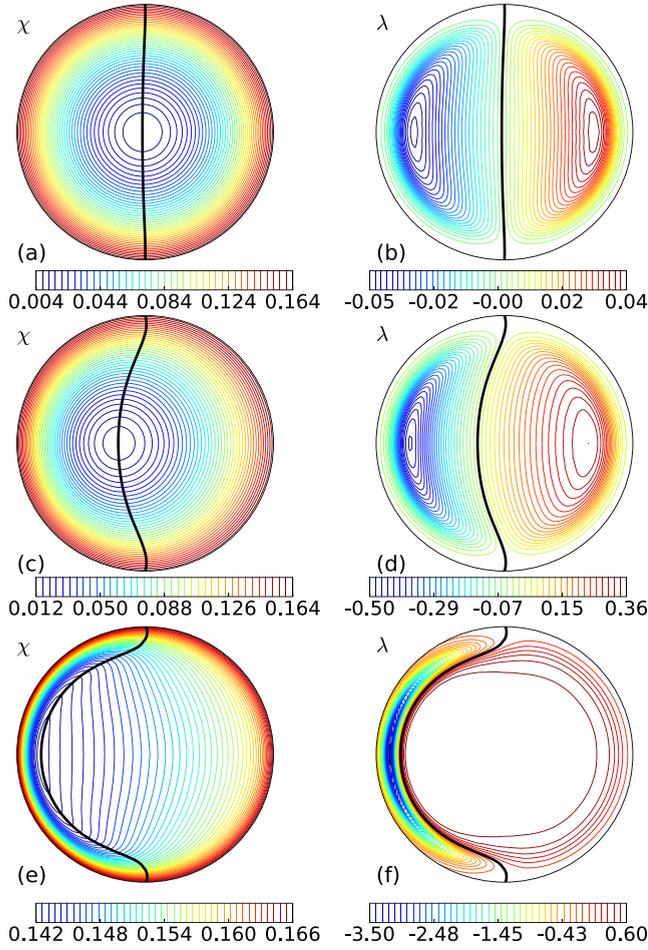

FIG. 4. Contours of the helical flux $\chi$ and $\lambda$ as the amplitude of the applied potential $\phi_0$ are increased. The two quantities, $\chi$ on the left and $\lambda$ on the right, are plotted, from top to bottom, for $\phi_0 = 2 \times 10^{-3}$, 0.02, and 0.6. $\lambda = 0$ contours (black) are also shown.

Neumann boundary conditions; (4) running with a loop voltage (back EMF); and last (5) performing a scan with respect to the Lundquist number $S$. The resistivity profile, issue (2), affects the magnitude of $\lambda$ in the reversal region. For the homogeneous Dirichlet conditions in issue (3), all components, $v_r$, $v_\theta$, $v_z$, are set to zero for all $(m, n)$, in contrast with the $E \times B$ boundary conditions imposed on $v_r$ in Sec. III. The homogeneous Neumann boundary conditions impose $\partial \mathbf{v}(1)/\partial r = 0$. Regarding issue (4), in the toroidal context, the loop voltage represents an inductive driving electric field in the same or opposite direction to the mean current provided by the helical electrodes. For straight plasma with electrodes at the ends, it corresponds to the back EMF of the circuit, depending on the load. In all cases, for the final saturated state driven with a large $\phi_0$ (or large $j_r$), we observe the same result of $q_0(r) \to 1$, with many of the other qualitative features remaining the same as in the nominal case.

## A. Specification of $j_r$ rather than $\phi_0$

An alternate way to impose the helical perturbation is by specifying the normal current density at the wall for $(1, 1)$ (and its complex conjugate), $j_r^{(1,1)}(r=1)$, while setting the tangential electric field to zero for all other pairs of $(m, n)$.

This approach can be used to benchmark the nominal case where an electrostatic potential for the $(1, 1)$ mode is specified at the wall, with all the other components $(m, n)$ of the tangential electric field again set to zero. The benchmark test entails recording $j_r^{(1,1)}(1)$ from a series of simulations with different values of $\phi_0$ once they have reached steady-state, then, programming this value of $j_r^{(1,1)}(1)$ as the boundary condition that simulates the helical drive, and finally recording the value of $\phi^{(1,1)}(1)$, which should match a particular $\phi_0$ from the set used. This also means that the resulting helical structures in the two cases must match in steady state. This check has been performed and indeed an agreement was found.

Figure 5 is the counterpart of Fig. 3 for the $j_r^{(1,1)}$ scan. As such, the same quantities, $1/q_0(0)$ (blue), $I_z$ (green, circles), $||\mathbf{v}_\perp||$ (red, arrows), and $r_O$ (cyan, squares), are plotted over a range in $j_r$ spanning $[10^{-4}, 0.13]$ that roughly corresponds to the same range used to scan $\phi_0$ in Sec. III C. The curves have the same qualitative shape as the four curves of Fig. 3. The inset figure in Fig. 5 shows $j_r^{(1,1)}$ vs. $\phi_0$, indicating a nonlinear dependence for $\phi_0 > 0.004$. This is not surprising since the steady state solution is strongly nonlinear in this range. This is also why $||\mathbf{v}_\perp||$ rises nonlinearly as a function of $j_r^{(1,1)}$ for $j_r^{(1,1)} \gtrsim 0.02$. Below this value, $1/q_0(0)$ and $I_z$ show a quadratic dependence as in Fig. 3, again because a quasilinear calculation of $j_z^{(0,0)}$ is accurate in this range.

It should not be concluded from this that if $\phi(r=1)$ is monochromatic, i.e., consists of a single Fourier harmonic $(m,n) = (1, 1)$ (plus of course $(m, n) = (-1, -1)$), then $j_r$ at the wall is also monochromatic. In fact, for $\phi_0$ specified for $(m, n) = (1, 1)$, with tangential electric field zero for all other $(m, n)$ (or for $j_r^{(1,1)}$ specified with tangential electric field zero for all other $(m, n)$) there is a spectrum of $j_r^{(m,m)}$ for $m > 1$ which becomes significantly broader as the system is driven harder, i.e., for large $\phi_0$ or large $j_r^{(1,1)}$. In fact, specification of $j_r^{(m,n)}$ for all $m$, $n$ is important for modeling helical electrodes that are localized in space, where one specifies $j_r(r = 1, u) = 0$ for the parts of the surface

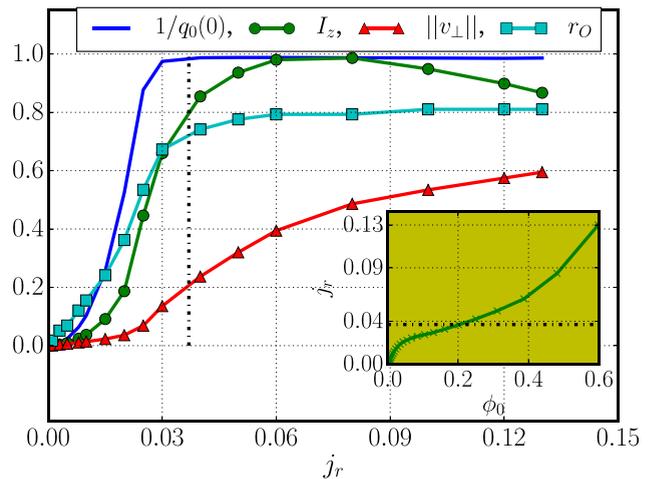

FIG. 5. Inverse $q_0(0)$ (blue), total axial current $I_z$ (green), $||\mathbf{v}_\perp||$ (red), and radial position of the O-point (cyan) as functions of the amplitude of the applied current density $j_r$. The inset figure shows $j_r$ vs. $\phi_0$, indicating a nonlinear dependence for $\phi_0 \gtrsim 0.004$. The dashed-dotted lines in both the main figure and the inset correspond approximately to the nominal case.

representing insulators. This issue will be treated in detail in a forthcoming publication.

### B. Resistivity profile

Since all the helical flux surfaces go through a thin layer near $r = 1$ (see Fig. 2(a)), we need to explore the importance of the resistivity profile in Eq. (4), which is quite peaked near $r = 1$ for the nominal case shown in Section III. To understand the sensitivity of the resistivity profile, the nominal case has been repeated with a uniform resistivity profile where $\eta_1 = \eta_0$ in Eq. (4) while keeping all of the remaining parameters the same as in Sec. III (for $\phi_0 = 0.2$). Figure 6 shows the same physical quantities $\chi$, $\lambda$, $g$, and $h$ as in Fig. 2. The flux surfaces and the $\lambda = 0$ contour are somewhat more helically distorted. The O-point has shifted outward to $r = 0.83$, similar to the position of the O-point ($r = 0.81$) for $\phi_0 = 0.6$ (strong drive) from Sec. III C. The $q_0(r)$ profiles from these two cases (Fig. 1) also closely resemble each other up to $r \approx 0.9$. The uniform-resistivity case bears more resemblance to the nominal case with $\phi_0 = 0.6$ than that with $\phi_0 = 0.2$, apparently because of its lower overall (volume-averaged) plasma resistivity, which effectively increases the helical drive. This agreement holds everywhere except for $r > 0.9$, where $q_0(r)$ blows up because $B_\theta^{(0,0)}$ reverses at $r = 0.98$, resulting in a slightly negative $I_z$. This reversal is not a fundamental property of having uniform resistivity, as it was discovered that $B_\theta^{(0,0)}$ remains small but positive when homogeneous Dirichlet flow boundary conditions (see Sec. IV C) are imposed. The general behavior appears to be $B_\theta^{(0,0)} \approx 0$, but not necessarily $B_\theta^{(0,0)} = 0$. Also note the greater volume is occupied by closed $g$ surfaces, indicating an increased amount of current that is not directly tied to the helical electrodes. This is the current that unequivocally flows down the length of the cylinder. The region of $\lambda < 0$ to the left has a much bigger $|\lambda|$ maximum that is located at the wall, as opposed to the case of Fig. 2(d). The enhancement of the $\lambda < 0$-region directly follows from the flux-surface average condition: since $\eta$ near the edge is now much smaller, $\lambda$ must compensate by becoming far more negative to satisfy $\langle \eta \lambda B^2 \rangle = \eta(0) \langle \lambda B^2 \rangle = 0$ near the edge. The $\lambda > 0$ region has a much flatter current density with a value just slightly above zero and the total current is slightly negative, as indicated above.

In spite of the aforementioned differences, the behavior observed for the case with a uniform plasma resistivity is qualitatively similar to that for the nominal case, although with a smaller $\phi_0$ for a comparable helical perturbation. The main qualitative differences are that for uniform resistivity the net current $I_z$ is slightly negative, as suggested by $\langle \lambda B^2 \rangle = 0$, and that the current inside closed $g$ surfaces is much larger.

### C. Alternative boundary conditions on the velocity

We first present results from a simulation with the first alternate specification of the velocity at the wall—described in Sec. II—which has all of its components at the wall equal to zero, in other words, a homogeneous Dirichlet boundary condition. All other parameters are the same as in the nominal case, in particular, $\phi_0 = 0.2$. The $\chi$, $\lambda$, and $h$ surfaces, displayed in Fig. 7, are not noticeably different from those in Fig. 2. The Lorentz force (not shown) is still localized to the wall, but is somewhat larger in the interior. The contours of $g$ are similar but do not have the small enclosed current surfaces $g = $ const. within a separatrix disconnected from the wall. The norm $||\mathbf{v}_\perp||$ decreases slightly from 0.20 to 0.18 (not shown) as can be expected since the boundary flows are set to zero.

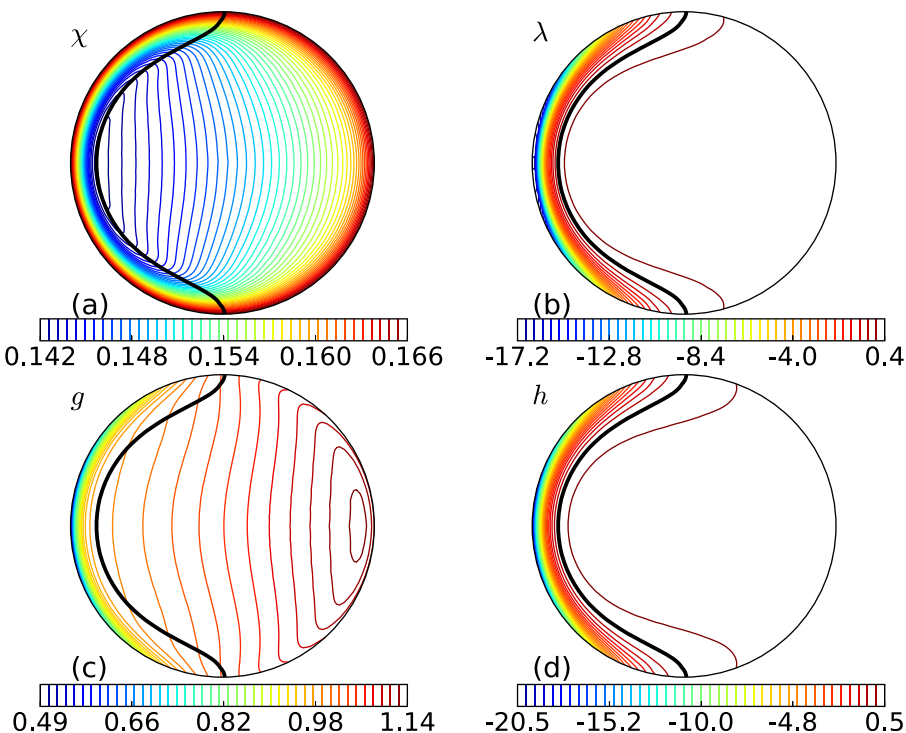

FIG. 6. The quantities $\chi$, $\lambda$, $g$, and $h$ from a simulation with a uniform $\eta$ profile and all of the other parameters the same as the nominal case ($\phi_0 = 0.2$). Note the enhanced distortion of the helical flux surfaces, greater volume occupied by closed $g$ surfaces, and larger shift of the O-point toward the wall. The parallel current density $\lambda$ is also more uniform in the $\lambda > 0$ region.

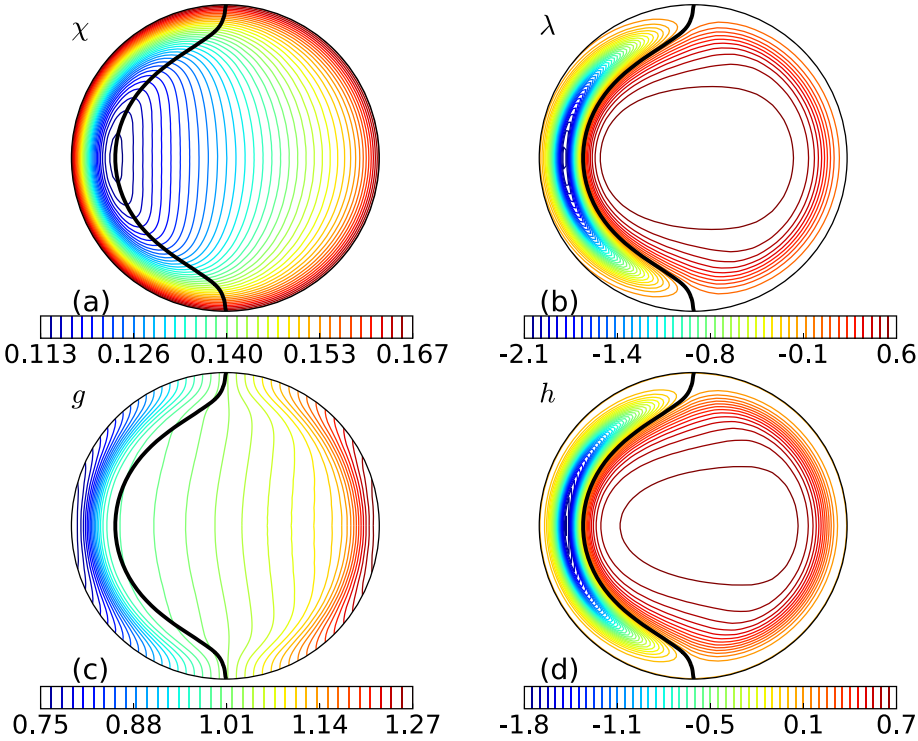

FIG. 7. The quantities $\chi$, $\lambda$, $g$, and $h$ for $\mathbf{v} = 0$ boundary conditions with all other parameters the same as in the nominal case ($\phi_0 = 0.2$) depicted in Fig. 2. While $\chi$, $\lambda$, and $h$ have undergone little change, the helical field ($g$) no longer shows any closed surfaces, indicating a sensitivity to the prescription of velocity at the wall.

Another alternative formulation of wall velocity that has been partially investigated imposes a homogeneous Neumann condition, $\partial \mathbf{v}(1)/\partial r = 0$. The most outstanding difference of this approach is the reappearance of the closed $g$ surfaces that occupy nearly one half of the domain (not shown here) for strong drive.

We conclude that while quantities such as $\chi$ and $\lambda$ do not depend noticeably on the choice of the velocity boundary condition, the helical field $g$ does, and therefore, so do the current streamlines. In fact, the existence of closed $g$ surfaces is quite sensitive to the choice of plasma flow at the wall. This effect will be studied in detail in a forthcoming publication.

### D. Back EMF or loop voltage $E_0$

In this subsection, we describe results with an applied loop voltage in the toroidal direction. One possible application is the study of toroidal plasmas, such as in RFPs and tokamaks, in which a helical electrode voltage might be applied in addition to the toroidal loop voltage. A loop voltage applied in the *opposite* direction to the mean plasma current density on axis provides a means to simulate the back EMF that should occur in a system of finite length if the axial current is drawn off through a load. A model with such an applied toroidal loop voltage in the resistive MHD cannot represent the effects of the sheaths at the electrodes at the ends, but it is a first step toward understanding how a finite length plasma with end electrodes responds to a load. Simulations with the endcap electrodes are outside the scope of the present study, but are related to flux core (or gun) spheromak studies.[21–24]

In the presence of loop voltage or back EMF, the Ohm's law in the Coulomb gauge in steady state is

$$E_0 \hat{\mathbf{z}} - \nabla \phi + \mathbf{v} \times \mathbf{B} = \eta \mathbf{j},$$

where $E_0 L$ is the loop voltage. With $\hat{\phi}(r, \theta, z) = \phi(r, u) - E_0 z$ we can write

$$-\nabla \hat{\phi} + \mathbf{v} \times \mathbf{B} = \eta \mathbf{j}.$$

Note that, although $\phi$ is single helicity, i.e., periodic in $z$, $\hat{\phi}$ is not. (In the Weyl gauge used in DEBS, the boundary condition on $A_z^{(0,0)}$ is $\partial A_z^{(0,0)}(r_w, t)/\partial t = -E_0$ and in steady state, we have $A_z^{(0,0)}(r, t) = A_z^{(0,0)}(r) - E_0 t$. As before, we have $\mathbf{A}(r, t) = \mathbf{A}_0(r) + t \dot{\mathbf{A}}_1(r)$, where $\dot{\mathbf{A}}_1$ incorporates $E_0$ as well as $\phi$; specifically, we have $\mathbf{E} = E_0 \hat{\mathbf{z}} - \nabla \phi = -\dot{\mathbf{A}}_1$.) The parallel component satisfies $E_0 B_z - \mathbf{B} \cdot \nabla \phi = \eta \lambda B^2$ and flux surface averaging yields

$$E_0 \langle B_z \rangle = \langle \eta \lambda B^2 \rangle \quad (15)$$

instead of Eq. (13). Also, the field line integration along the $O$-point gives

$$E_0 B_z = \eta \lambda B^2, \quad (16)$$

with all quantities constant since they are evaluated at the $O$-point, as in the arguments in Sec. III B. Thus, unlike the cases with $E_0 = 0$ discussed in that section, the value of $\lambda$ at the $O$-point is negative for $E_0 < 0$.

We have performed simulations using the same parameters as in the nominal case, except $E_0 < 0$. When $E_0 = -0.01$ ($|E_0 L| \approx |\phi_0|$), $B_\theta^{(0,0)}$ at the wall, and hence the total current $I_z = 2\pi \int j_z r dr$, reverse. The $q_0(r)$ profile remains flat while $q_0(0)$ increases slightly as $E_0$ becomes more negative. The current $I_z$ is observed to decrease linearly with $E_0$. Figure 8 shows the helical flux with the curve $\lambda = 0$ for $E_0 = -0.01$,

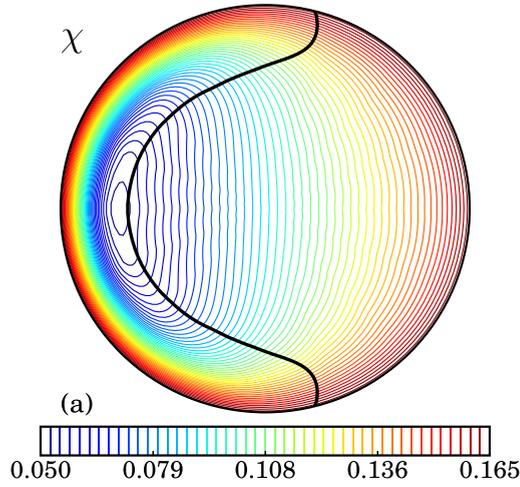

FIG. 8. Surfaces of constant helical flux $\chi$ for a case with all parameters as in the nominal case, except with $E_0 = -0.01$. The curve $\lambda = 0$ has moved to the right, expanding the region with $\lambda < 0$, consistent with a decrease in the total current $I_z$. Note that the $O$-point is in the region $\lambda < 0$, as expected from Eqs. (15) and (16).

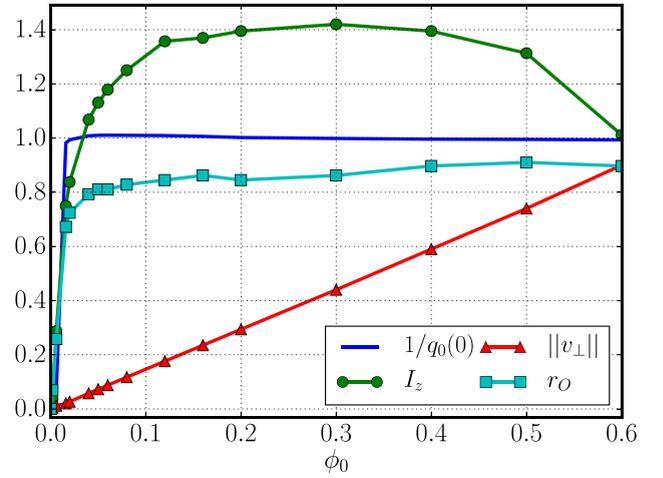

FIG. 9. Inverse $q_0(0)$ (blue), total axial current $I_z$ (green), $||\mathbf{v}_\perp||$ (red), and radial position of the $O$-point (cyan) as functions of the amplitude of the applied electrostatic potential $\phi_0$ for $S = 1000$ and $Pr = 100$ (such that $Re$ remains fixed.)

with all other parameters as in the nominal case (Fig. 2.) In accordance with the conclusions above, the curve $\lambda = 0$ in Fig. 8 no longer intersects the $O$-point in the $\chi$ surfaces. In fact, $\lambda_O < 0$ and the curve $\lambda = 0$ (as well as the nearby $j_z = 0$ curve) move to the right as $E_0$ decreases. That is, the $\lambda < 0$ region grows, leading to more cancellation of the positive and negative current. The position of the $O$-point also moves to the right, but by a lesser amount. For cases in which $E_0$ is positive, the total current $I_z$ increases, and the surface $\lambda = 0$ (approximately $j_z = 0$) moves to the left (with the $O$-point position moving to the left a lesser amount), decreasing the cancellation of the positive current density by the negative current density.

### E. Lundquist number scan

Another topic of interest is the robustness of the aforementioned characteristics of the nominal state to hotter plasmas. Since the plasma temperature is not evolved for the present study, a "hotter" plasma is implemented by scaling down the resistivity (scaling up the Lundquist number $S$). For achieving this, additional resistive MHD simulations have been run at $S = 1000$ and $S = 5000$. For $S = 1000$, we consider two cases: one with a fixed viscosity (i.e., fixed Reynolds number $Re$); and the other with a fixed magnetic Prandtl number $Pr = S/Re$. The latter case amounts to increasing $Re$ by a factor of 10 from the value used in the $S = 100$ simulations. A full scan has been carried out for the former case with a partial scan for the latter. All of the remaining parameters are kept the same as in the nominal case.

The main characteristics remain the same as functions of $\phi_0$ in the moderate-to-strongly driven regime, as indicated by Fig. 9. When $\phi_0$ is sufficiently large, $q_0(0)$ approaches unity from above, and remains close to unity over a wide range in $\phi_0$. Note that the "knee" in the $1/q_0(0)$ vs $\phi_0$ curve is steeper than that for the $S = 100$ scan. The typical flat $q_0(r) \gtrsim 1.0$ profiles once again appear inside the plasma, as indicated by Fig. 10. These profiles are even flatter than those for $S = 100$, displayed in Fig. 1. Again, the radial position of the $O$-point, shown in Fig. 9, monotonically shifts outward toward $r = r_w$, while $||\mathbf{v}_\perp||$ linearly appears to rise as $\phi_0$ is increased. The plasma current grows with $\phi_0$ until it reaches a critical point, beyond which it begins to decrease somewhat, as in Figs. 3 and 5. The maximum value of $I_z$ is approximately 40% greater than that of $S = 100$ (Fig. 2), although the current density $j_z^{(0,0)}(r = 0) \sim 2B_z^{(0,0)}/Rq_0(0) \to 1$ remains about the same. For $\phi_0 \gtrsim 0.6$ where $||\mathbf{v}_\perp||/v_A = 0.6$, the time-independent state with a flat $q_0(r)$ profile transitions into a time-varying state. These findings are consistent with observations from Secs. III and IV A that the steady state is lost when $||\mathbf{v}_\perp||/v_A$ becomes of order unity. However, note the wider range of $\phi_0$ in Fig. 9 over which $q_0(0) \approx 1.0$. The results of the second scan with $S = 1000$ at a fixed $Pr = 10$ are nearly identical to those of the $S = 1000$ and $Pr = 100$ (fixed $Re$).

Figure 11 shows $\chi$, $\lambda$, and $g$ for two different values of $\phi_0$; 0.02 (left column) and 0.2 (right column) at $S = 1000$

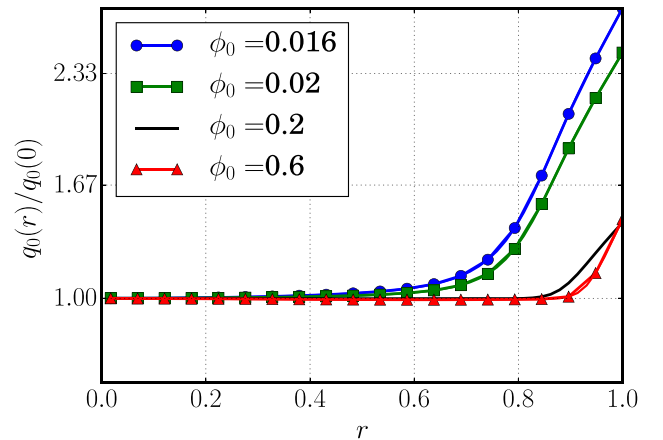

FIG. 10. The mean field profile $q_0(r)$ normalized by $q_0(r = 0)$ for the case with $S = 1000$ and $Pr = 100$ driven with four different values of the applied potential: $\phi_0 = 0.016$ (blue circles), 0.02 (green squares), 0.2 (solid black), and 0.6 (red triangles) for a $(m,n) = (1, 1)$ electrostatic perturbation.

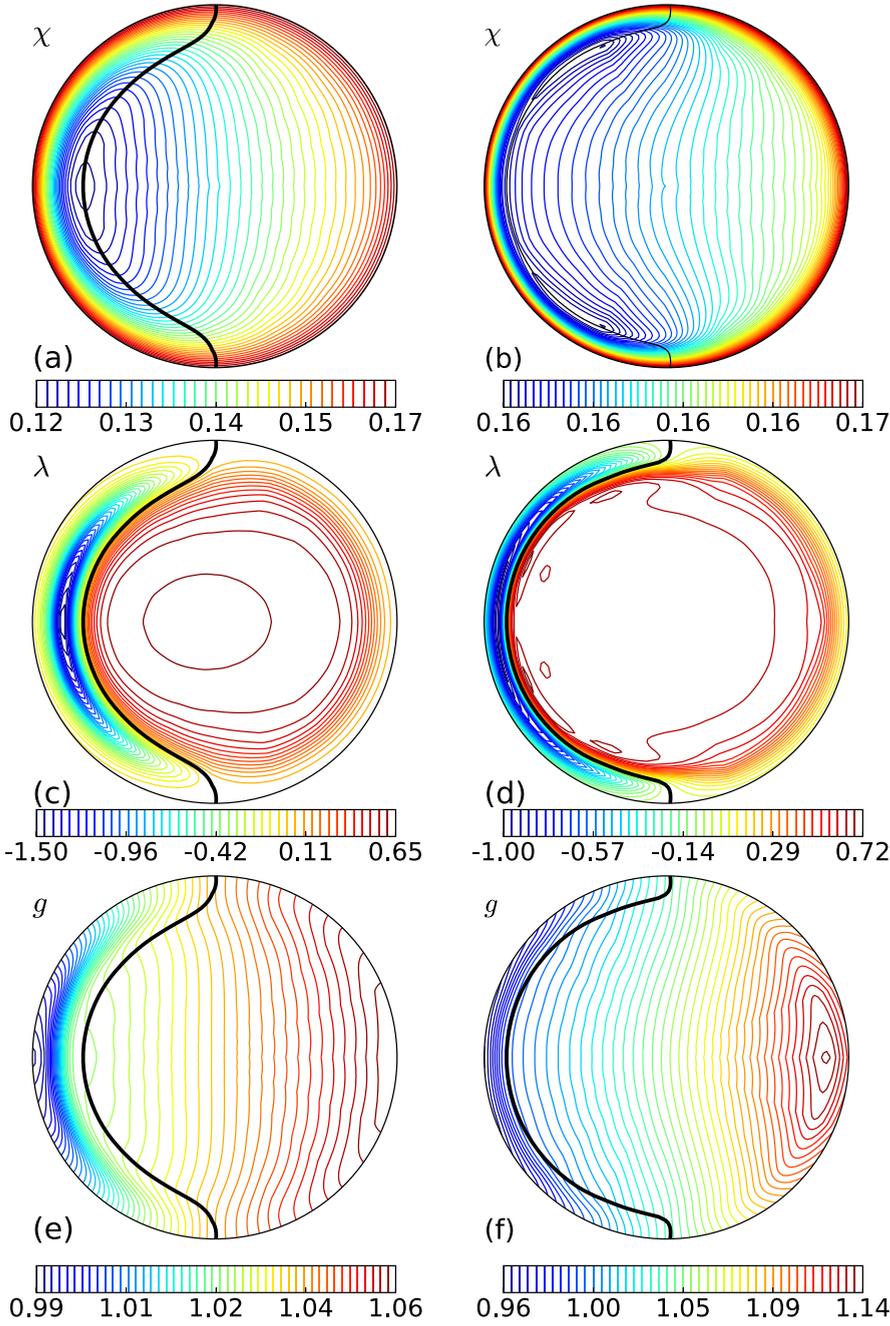

FIG. 11. Contours of the salient physical quantities $\chi$ (top row), $\lambda$ (middle row), and the helical field $g$ (bottom row) from two simulations with $S=1000$ and $Pr=100$ (fixed $Re$ number). Left column corresponds to a moderately driven case with $\phi_0=0.02$, and the right column to a strongly driven case with $\phi_0=0.2$. In the strong-drive regime, the $\chi$ surfaces transition to a configuration from a single $O$-point to one with two $O$-points and a single $X$-point that connects them, all having $\lambda=0$.

and $Pr=100$ (fixed $Re$). There is a critical value of the potential, $\phi_c \approx 0.06$, beyond which the structure in $\chi$ undergoes a noticeable change and exhibits two $O$-points and an $X$-point, all of which are intersected by the $\lambda=0$ curve (Fig. 11(b)). These additional structures give rise to spikes in $\lambda$ observed in Fig. 11(d). This bifurcation in $\chi$ surfaces from one $O$-point to two $O$-points and an $X$-point is suggestive of a tearing instability because of the elongation of the $\chi$ surfaces near the $O$-point. The region with $\lambda>0$ is flatter and extends over an area up to $r=0.8$ that is centered at the origin. In agreement with the trends of Sec. III C, closed $g$ surfaces appear as $\phi_0$ increases, with the current in the closed $g$ surfaces increasing with $\phi_0$.

The results from the $S=5000$ scan (with $Pr=100$) are very similar to those from $S=1000$ scan. The existing trends become more pronounced with increasing $S$. The "knee" in the $1/q_0(0)$ curve is steeper and shifted to $\phi_0=0.002$. The steady state is again lost at $\phi_0 \approx 0.6$. Note that a time-dependent state develops at nearly the same value of $\phi_0$ (and $||\mathbf{v}_\perp||/v_A$) for all three values of $S$. For the same strength of the helical potential, the region of flat $\lambda>0$ spans even a bigger area than that at $S=1000$. The two $O$-points for $S=5000$ (not shown) are poloidally farther separated than observed for $S=1000$ (Fig. 11(b)). The transition from a single $O$-point configuration to one with two $O$-points and an $X$-point occurs at a lower value $\phi_0=\phi_c \approx 0.01$.

## V. MAGNETIC HELICITY

The preceding considerations lead to the closely related issue of the role of magnetic helicity in these states. The relative magnetic helicity is given by[25]

$$K = \int_V (\mathbf{A} - A_{zw}^{(0,0)}\hat{\mathbf{z}}) \cdot \mathbf{B} d^3x, \quad (17)$$

where $A_{zw}^{(0,0)}$ is the value at $r = r_w$. See also Ref. 26. The general gauge-invariant expression for relative helicity,[25] $\int (\mathbf{A} + \mathbf{A}') \cdot (\mathbf{B} - \mathbf{B}') d^3x$, where $\mathbf{A}'$, $\mathbf{B}'$ are reference (vacuum) quantities, takes the form in Eq. (17) since $\nabla \times (A_{zw}^{(0,0)}\hat{\mathbf{z}}) = 0$. The wall term $A_{zw}^{(0,0)}$ is constant in time for $E_0 = 0$. According to Refs. 25–27, for $E_0 = 0$ the magnetic helicity injection rate is given by $\dot{K}_{inj} = -2\int_S \phi B_n dS$, which is zero in this case because of the boundary condition $B_n = B_r = 0$. Thus, the total helicity injection rate from the helical electrostatic perturbation for $E_0 = 0$ is zero for all times. On the other hand, the rate of change of the total helicity due to resistive diffusion in the plasma is

$$\dot{K}_p = -2\int_V \eta \mathbf{j} \cdot \mathbf{B} d^3x = -2\int_V \eta \lambda B^2 d^3x, \quad (18)$$

where $V$ represents the total volume. We have argued in Sec. III B that $\langle \eta \lambda B^2 \rangle$ is zero on any flux surface in steady state. This quantity equals $(d/dV)\int \eta \lambda B^2 d^3x$ (see the Appendix), leading to $\dot{K}_p = 0$ in steady state. $V = V(\chi)$ is the volume inside the flux surface. We have performed the flux surface averages of the computational results in this nominal case, showing that the expression for $\dot{K}_p$ in Eq. (18) is indeed very small in steady state.

However, this and $\dot{K}_{inj} = 0$ do not imply that the helicity in the steady state solution is zero, because $\dot{K}_p$ can be non-zero in the transient period. In fact, $\dot{K}_p$ is positive in the transient, leading to positive helicity in the steady state in spite of zero helicity injection, $\dot{K}_{inj} = 0$. Indeed, the magnetic helicity in the steady state for the nominal case is $K = 6.6$. As a check, we have computed $\int_0^t \dot{K}_p(t')dt'$ over the transient and found that it agrees with this steady state value of $K$. In the steady state, helicity is generated where $\lambda < 0$ and dissipated where $\lambda > 0$. The possibility of generating helicity with $\dot{K}_{inj} = 0$ in regions with $\lambda < 0$ has been pointed out in a different context in Ref. 28. Figure 12 shows the helicity as a function of time for the nominal case $E_0 = 0$ of Sec. III (solid black), exhibiting a clear increase during the transient period ($t/\tau_A \lesssim 15$). Also shown is the integral of $\dot{K}_p$ in time (blue crosses), showing good agreement.

Equation (17) holds regardless of the value of the back EMF (or loop voltage) $E_0$. For $E_0 \neq 0$, $A_{zw}^{(0,0)}$ has a term proportional to $E_0 t$ and the helicity injection rate is given by $\dot{K}_{inj} = -2\int_S \phi B_n dS + 2E_0 L \Phi_w$, where $\Phi_w$ is the total flux $\int B_z r dr d\theta$. Again, $B_r = 0$ at the wall implies

$$\dot{K}_{inj} = 2E_0 L \Phi_w \quad (19)$$

and again $\dot{K}_p = -2\int_V \eta \lambda B^2 d^3x$. For $E_0 < 0$, $\dot{K}_{inj}$ is negative, implying that back EMF takes helicity out of the system. Figure 12 also shows results for three negative values of $E_0$. It is seen from this figure that negative $E_0$ results in $\dot{K} < 0$ initially. However, due to the contribution by the $\dot{K}_p$ term, the helicity saturates to a final steady state value ($\dot{K} = 0$), which can be positive or negative. Indeed, Eq. (15) implies $E_0 \int B_z d^3x = \int \eta \lambda B^2 d^3x$, showing $\dot{K}_p + \dot{K}_{inj} = 0$. The results in Fig. 12 show that the steady state helicity decreases linearly as a function of $E_0$ and becomes negative at $E_0 \approx -0.018$.

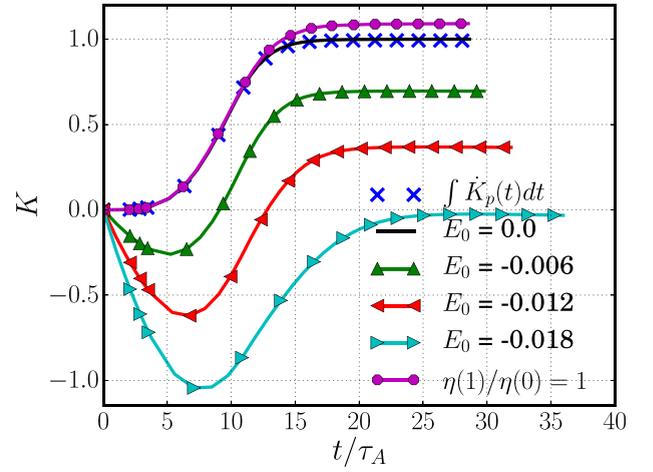

FIG. 12. Magnetic helicity as a function of time for $E_0 = 0, -0.006, -0.012, -0.018$. The case with $E_0 = 0$ (solid black) corresponds to the nominal case, which uses no back EMF. For the first case, the values of $\int \dot{K}_p dt$ are also included, indicated by the blue crosses, and show good agreement with $K(t)$ (solid black). The purple line with circles corresponds to the case with a uniform resistivity for $E_0 = 0$ (see Sec. IV B). It reaches a higher steady-state helicity because there is less overall helicity dissipation, since the effective (volume-averaged) $\eta(r)$ is smaller than that for the nominal case. All values of helicity are normalized to the steady state value for the nominal case ($E_0 = 0$).

If we assume the gauge condition $A_r = 0$, we obtain from Eq. (17)

$$K = \int \left( \frac{\psi}{r} \frac{\partial \Phi}{\partial r} - \frac{\Phi}{r} \frac{\partial \psi}{\partial r} \right) r dr d\theta dz, \quad (20)$$

$$= 2\int \psi B_z r dr d\theta dz, \quad (21)$$

where $\psi = A_z - A_z^{(0,0)}$ and $\Phi = rA_\theta$, and we have used $\psi(r=1) = 0$, $\Phi(r=0) = 0$ to carry out the integration by parts in obtaining Eq. (21). Writing $\chi = \psi + \Phi/R$ for $m = n = 1$, we find the two equivalent forms

$$K = 2\int \chi B_z d^3x - 4\pi^2 \Phi^2(r=1), \quad (22)$$

$$K = 2\int \left( \chi - \frac{\Phi(r=1)}{2R} \right) B_z d^3x. \quad (23)$$

This form shows that $K$ can be negative, as it is for $E_0 = -0.018$, without $\chi$ changing sign. The form in Eq. (21) shows that $K$ is the linkage between the poloidal flux $\psi$ and the toroidal flux $\Phi$. The forms in Eqs. (22) and (23) show that in terms of linkage between $\chi$ and $\Phi$, there is an extra factor that arises because of helical geometry.

For $E_0 > 0$, $\dot{K}_{inj}$ is positive; this sign of driving loop voltage drives the current in the $+z$ direction, with $q_0$ positive. Again, we have compared $K(t)$ with $\int (\dot{K}_p + \dot{K}_{inj}) dt$ and found them to be in agreement.

## VI. APPLICATIONS

Our model, which is periodic in $z$, can be used as a rough guide for understanding a plasma with helical and endcap electrodes for DC transformer applications.[29] The nominal case studied in Sec. III has electrostatic potential $\phi$ (in the Coulomb gauge) periodic in the $z$-direction with zero back EMF ($E_0 = 0$).

Recall the large degree of cancellation in $I_z$ due to the bipolar structure of $j_z$. This suggests slotting the endcap electrodes with a gap filled with insulating material.[30] A reasonable position for the gap is along the curve $\lambda = 0$ (nearly the same as $j_z = 0$), as shown, for example, in Fig. 2(b). If the endcaps at both ends are slotted, we can consider the two segments with $j_z > 0$ to be connected to a circuit and the two segments with $j_z < 0$ to be connected to another circuit, and these two circuits can be added either in series or in parallel. In the former case, the voltages would add, and in the latter case, the currents would add. Another possibility is slotting one of the endcap electrodes and not the other so that one sign of current density can flow into the unslotted electrode and back out of the same electrode, with the two segments of the slotted electrode connected to a circuit. In this configuration, the two halves of the plasma with opposite signs of $j_z$ are effectively connected in series. The periodic simulations in this paper can only be suggestive for such schemes involving slotted endcap electrodes. Results from simulations with finite length plasmas await further publication.

Another possible application for this scheme of driving a plasma with helical electrodes is to tailor the current density profile in a tokamak, or possibly in an RFP. For the tokamak, the present results in a periodic domain can be applied directly, although toroidal effects are not included. Without a toroidal loop voltage, a tokamak can still have the current provided by the bootstrap current,[7] and there may be neutral beam driven (Ohkawa) current also.[7] However, the bootstrap current density is hollow; in fact, the bootstrap current at the magnetic axis is zero.[7] The current density supplied by the electrostatic scheme of this paper provides current density that is peaked at the center (not necessarily peaked at the magnetic axis but still positive) and should cancel the bootstrap current near the edge, and may allow for a peaked or only slightly hollow current density profile.

## VII. SUMMARY

This paper features an initial study of a $\beta = 0$ plasma in periodic cylindrical geometry driven by a helical electrostatic potential applied at its edge. The initial vacuum field $B_z = B_0$ is uniform in space. The perturbation at the wall $r = r_w$ is of the form $\phi_0 e^{im\theta + ikz} = \phi_0 e^{i\theta - i\zeta}$, where $k = -n/R$ and $\zeta$ is the toroidal angle $\zeta = z/R$. Here, we have taken $m = 1$ and $n = 1$. For weak $\phi_0$ the time asymptotic state consists of a very small $m = n = 1$ perturbation treated well by a linear response. For larger $\phi_0$ the time asymptotic state is still single helicity, i.e., a broader spectrum with $m/n = 1$. This state has highly distorted helical flux surfaces (surfaces of $\chi = mA_z - krA_\theta$), like those of a full Kadomtsev reconnection[6–8] or a SHAx state in an RFP.[9,10]

The time asymptotic state for a large $\phi_0$ is well characterized by nearly Alfvénic helical flows and a flat quasilinear safety factor profile $q_0 = rB_\theta^{(0,0)}/RB_z^{(0,0)} \gtrsim m/n = 1$ except near the plasma edge. Consistent with this, there is a large area with $\mathbf{k} \cdot \mathbf{B} \approx 0$, which allows this strong distortion of the $\chi$ surfaces for a relatively small magnetic perturbation. This $q_0(r) \gtrsim 1$ is a consequence of a quasilinearly generated mean field $B_\theta^{(0,0)}$ and slightly paramagnetic $B_z^{(0,0)}$. The current density lines, on the surfaces of constant helical field, $g = mB_z - krB_\theta$, follow the $\chi$ surfaces fairly well in the nearly force-free interior, but deviate near the edge where they cross the magnetic surfaces to go into and out of the helical electrodes. In the strongly driven regime, closed $g$ surfaces can appear, indicating a flow of current along the axis of the cylinder disconnected from the helical electrodes. However, our results indicate a dependence of the presence of closed $g$ surfaces on the velocity boundary conditions. For even larger $\phi_0$ the time asymptotic state is no longer a steady state.

Another distinct characteristic of this state comes from the flux surface average of $\eta \mathbf{j} \cdot \mathbf{B} = \eta \lambda B^2$, which is equal to zero, implying that $\lambda$ consists of Pfirsch-Schlüter currents, with $\mathbf{j}_\perp$ balanced by inertial and viscous stresses rather than pressure. This also implies that $\lambda = 0$ on the magnetic axis (O-line) in steady state. We have shown how the flux surface average arguments apply to the evaluation of the magnetic helicity. We have also noted that $j_z$ and $\lambda$ reverse on approximately the same surface, so that reversal of $j_z$ is closely related to the flux surface arguments relating to $\lambda$. For application to a plasma of finite length with electrodes at the ends, this change of sign of $j_z$ suggests the possibility of slotting the end electrodes of the device.

We have investigated the sensitivity of this state to various changes. These include: (1) Specifying the current density $j_r$ at the helical electrodes rather than the potential $\phi_0$; (2) Varying the profile of the resistivity from one peaked near the wall to a uniform profile; (3) Changing the boundary conditions on the radial velocity at $r = r_w$ from $E \times B$ flow to $v_r = 0$, thereby $\mathbf{v}(r_w) = 0$, and to $\partial \mathbf{v}(r_w)/\partial r = 0$; (4) Including back EMF or loop voltage and studying the magnetic helicity; and lastly (5) varying the Lundquist number $S$. For items (1), (2) (and mostly (5)), the qualitative features of the nominal state remain unchanged. While most features underwent little change for item (3), it was found that the appearance of closed $g$ surfaces is quite sensitive to the choice of flow boundary condition. A further study of this issue is deferred to a future publication. In the case of item (4), the presence of back EMF, $E_0 L$, where $L = 2\pi R$, implies that the flux surface average of $\eta \lambda B^2$ is balanced by the flux surface average of $E_0 B_z$, so that for $E_0 < 0$ the value of $\lambda$ at the O-point is negative. For item (5), raising $S$ makes the defining characteristics more pronounced. For example, the $q_0(r)$ profile becomes flatter and the range of $\phi_0$ with a flat $q_0(r) \gtrsim m/n = 1$ expands.

Section V includes a discussion of magnetic helicity for the transient period and the time-asymptotic state. The rate of change of helicity $\dot K$ is zero for the steady-state, as expected. Although $B_n = 0$, implying that the helicity injection rate is zero, $\dot K \neq 0$ during the transient period, which results in a finite non-zero final magnetic helicity. What is commonly thought of strictly as the helicity dissipation term is responsible, here, for generating positive magnetic helicity where $\lambda < 0$ and vice-versa. The back EMF results initially in $\dot{K} < 0$. If $|E_0|$ is large enough ($|E_0|L \gtrsim |\phi_0|$), the helicity for the time-asymptotic state can be negative.

In Sec. VI, we have described two possible applications of this unique steady state. These are (1) the development of direct current (DC) electrical transformers and (2) the possibility of tailoring the current density profile in a tokamak or possibly a RFP.

Future work will investigate steady-state solutions of the helical drive with harmonics $(m, n)$ other than $(1, 1)$ and the dependence of the above characteristics on the aspect ratio for the periodic cylinder. However, our main focus will be on describing the physics in a cylinder of finite length where the ends of the cylinder could be perfectly or partially conducting. The former results in the line-tying of the axial magnetic field for perfectly conducting ends. New diagnostics will have to be implemented because of the lack of helical symmetry in finite-length geometries.

## ACKNOWLEDGMENTS

We thank the remaining Tibbar staff members: Juan Fernandez, William Gibson, Neal Martin, Aaron McEvoy, Keith Moser, and Liviu Popa-Simil as well as Carl Sovinec from UW Madison for their input and valuable discussions. We also thank the anonymous referees for their suggestions. This research was supported by funding from the ARPA-E agency of the Department of Energy under Grant No. DE-AR000067. The Python software was used to generate the graphics.

## APPENDIX: ON THE HELICAL SYMMETRY DIAGNOSTICS AND ENERGY CONSERVATION

In helical geometry, the magnetic field is represented by

$$\mathbf{B} = f(\nabla \chi \times \boldsymbol{\sigma} + g\boldsymbol{\sigma}), \tag{A1}$$

where $f(r) = 1/r|\mathbf{k}|^2$, $\mathbf{k} = \nabla u$, and $\boldsymbol{\sigma} = \hat{\mathbf{r}} \times \mathbf{k}$. The current density, also being solenoidal, has the same general form as **B**, namely, $\mathbf{j} = f(\nabla Q \times \boldsymbol{\sigma} + h\boldsymbol{\sigma})$, and $\nabla \times \mathbf{B} = \mathbf{j}$ implies $Q = g$ or

$$\mathbf{j} = f(\nabla g \times \boldsymbol{\sigma} + h\boldsymbol{\sigma}). \tag{A2}$$

In these equations, we have $\chi = r\boldsymbol{\sigma} \cdot \mathbf{A}$ and $g = r\boldsymbol{\sigma} \cdot \mathbf{B}$, the helical flux and the helical field, respectively. Furthermore, $B_r = (1/r)\partial \chi/\partial u$ and $\mathbf{k} \cdot \mathbf{B} = -(1/r)\partial \chi/\partial r$. These relations, and similar relations between current density and $g$, show

$$\mathbf{B} \cdot \nabla \chi = 0, \quad \mathbf{j} \cdot \nabla g = 0, \tag{A3}$$

so that magnetic field lines lie on $\chi = $ const. surfaces and current density streamlines lie on $g = $ const. surfaces. Similarly, we have $j_r = (1/r)\partial g/\partial u$, $\mathbf{k} \cdot \mathbf{j} = -(1/r)\partial g/\partial r$ and $\boldsymbol{\sigma} \cdot \mathbf{j} = h/r$.

An elliptic or $O$-line, i.e., a maximum or minimum of $\chi$ as in Fig. 2(a), is possible in Ohmic equilibrium with zero current density because of the helical nature of the perturbations. Specifically, if $\boldsymbol{\sigma} \cdot \nabla \times \boldsymbol{\sigma} \propto mk$ is not zero, there can be an extremum of $\chi$ with zero current density.[2] With either $m = 0$ or $k = 0$ an $O$-line requires local current density, which will decay due to resistivity.

For a force-free plasma, with $\mathbf{j} \times \mathbf{B} = 0$, it can be shown that $g = g(\chi)$ and $h = h(\chi)$ as well as $h(\chi) = g(\chi)g'(\chi)$ and $\lambda(\chi) = g'(\chi)$. However, because the plasma is not exactly force-free, the contours of $\chi, g,$ and $h$ shown in Figs. 2, 7, and 11 differ significantly. The Lorentz force ($\mathbf{j} \times \mathbf{B}$) is fairly strong near the wall and smaller in the interior. It is balanced by the inertia and/or the viscous stresses rather than pressure gradient.

An important tool for diagnosing plasma flows, particularly, the perpendicular flows, is $\phi$, the electrostatic potential (in the Coulomb gauge.) From $\mathbf{E} = -\nabla \phi$, we have

$$\nabla^2 \phi = -\nabla \cdot \mathbf{E}. \tag{A4}$$

The quantity $\nabla \cdot \mathbf{E} = \rho_q$ is the charge density, which does not enter directly into the MHD equations because of quasi-neutrality, but Eq. (A4) can still be solved for $\phi$. The boundary conditions are inhomogeneous Dirichlet, namely,

$$\phi(1, m\theta + kz) = \phi_0 e^{im\theta + ikz} = \phi_0 e^{iu} \tag{A5}$$

for $m = n = 1$ ($k = -n/R$), and zero for $(m, n) \neq 1$, which is exactly the helical potential applied at the boundary, according to Eqs. (5)–(7). For **E** and $\rho_q$ in Fourier representation, e.g., $\rho_q = \sum_l \rho_{ql}(r)e^{il(m\theta+kz)}$ ($m = n = 1$) we solve

$$\frac{1}{r}\frac{\partial}{\partial r}\left(r\frac{\partial \phi_l}{\partial r}\right) - l^2\left(\frac{m^2}{r^2} + k^2\right)\phi_l = -\rho_{ql}$$

for each value of $l$.

The $E \times B$ velocity is given by

$$\mathbf{v}_{E \times B} = \frac{\hat{\mathbf{b}} \times \nabla \phi}{B}, \tag{A6}$$

so that $\mathbf{v}_{E \times B} \cdot \nabla \phi = 0$. Its divergence is given by $\nabla \cdot \mathbf{v}_{E \times B} = \nabla(1/B) \cdot \hat{\mathbf{b}} \times \nabla \phi + (1/B)\nabla \phi \cdot \nabla \times \hat{\mathbf{b}}$. For large aspect ratio with $q \sim 1$, $B$ is nearly constant and $\hat{\mathbf{b}} \approx \hat{\mathbf{z}}$, showing that $\nabla \cdot \mathbf{v}_{E \times B}$ is small, away from the wall. As discussed in Sec. III A, the total perpendicular velocity $\mathbf{v}_\perp$ has a correction to $\mathbf{v}_{E \times B}$ proportional to $\eta \mathbf{j} \times \mathbf{B}$. This correction is fairly small in the interior and appreciable near the boundary, so that $\mathbf{v}_\perp$ aligns with the $\phi$ surfaces in the interior, as shown in Fig. 2(e). The contribution due to $\eta \mathbf{j} \times \mathbf{B}$ and therefore the divergence, can be appreciable near the boundary. This issue will be studied further in a future publication.

Flux surface averages are discussed for these helically symmetric steady state solutions in the body of the paper. We define a flux surface average in terms of the volume average within a region $R(\chi)$ labeled by $\chi = $ const., with $\chi'(r, u) \leq \chi$. The volume integral of a quantity $f$ and the volume $V(\chi)$ are defined by

$$F(\chi) = \int_{R(\chi)} f(\mathbf{x}) d^3x, \quad (A7)$$

with

$$V(\chi) = \int_{R(\chi)} d^3x, \quad (A8)$$

and the flux surface average of an arbitrary function $f(r, u)$ is given by

$$\langle f \rangle_\chi = \frac{dF(\chi)/d\chi}{dV(\chi)/d\chi} = \frac{dF}{dV}. \quad (A9)$$

Also, $F$ can be written as $F(\chi) = \int_{R(\chi)} f \, d\chi' dS/|\nabla\chi|$, so we have

$$\langle f \rangle_\chi = \frac{\int_{S(\chi)} f \, dS/|\nabla\chi|}{\int_{S(\chi)} dS/|\nabla\chi|}, \quad (A10)$$

where $S(\chi)$ is the surface of $R(\chi)$. The volume integral $\int_{R(\chi)} \mathbf{B} \cdot \nabla f d^3x$ equals $\int_{S(\chi)} f B_n dS$, which is zero since $B_n = 0$ on $S(\chi)$. Therefore, it is clear that Eq. (A10) defines the exact weighting in the flux surface average that gives $\langle \mathbf{B} \cdot \nabla f \rangle_\chi = 0$. In the computations in this paper, we perform these flux surface averages by obtaining $F(\chi)$ and $V(\chi)$ and using Eq. (A9).

Next, we review energy conservation in ideal MHD under the assumption of constant density ($\rho = \rho_0$). The energy of a zero-pressure system in a volume $V$ that is bounded by a surface $S$ is

$$E = \int_V \left( \frac{\rho_0 \mathbf{v}^2 + \mathbf{B}^2}{2} \right) d^3x. \quad (A11)$$

Then, the energy conservation in ideal MHD takes one of two forms assuming $\rho = \rho_0 = $ const. and $B_n = 0$ on $S$:

$$\frac{dE}{dt} = -\int_V \rho_0 \mathbf{v} \cdot \nabla \left( \frac{\mathbf{v}^2}{2} \right) - \int_S \hat{\mathbf{n}} \cdot \mathbf{E} \times \mathbf{B} dS \quad (A12)$$

or

$$\frac{dE}{dt} = -\int_V \rho_0 \mathbf{v} \cdot \nabla \left( \frac{\mathbf{v}^2}{2} \right) - \int_S B^2 v_n dS. \quad (A13)$$

That is, under these conditions, $v_n$ equals $\hat{\mathbf{n}} \cdot (\mathbf{E} \times \mathbf{B})/B^2$. The second term in either expression is the Poynting flux. The first term is present because of the condition $\rho_0 = $ const., leading to a violation of energy conservation. However, the magnitude of this term is approximately $\rho_0 v^3 / r_w \sim (v/v_A)^3$. Thus, the energy loss is insignificant except for the most strongly driven cases where $v/v_A \sim 1$. For example, for the nominal case of Sec. III A, $(v/v_A)^3 \sim 0.008$ and the violation of energy conservation is measured to be less than 1%.